# Ensemble model for pre-discharge icd10 coding prediction

Yassien Shaalan[1],PhD, Alexander Dokumentov[1], PhD, Piyapong Khumrin[2], MD,PhD, Krit Khwanngern[3], MD, Anawat Wisetborisut[4], MD, Thanakom Hatsadeang[5], Nattapat Karaket[6], Witthawin Achariyaviriya[7], Sansanee Auephanwiriyakul[8], Nipon Theera-Umpon[9], PhD, Terence Siganakis[1], MSc

**Abstract.** The translation of medical diagnosis to clinical coding has wide range of applications in billing, aetiology analysis, and auditing. Currently, coding is a manual effort while the automation of such task is not straight forward. Among the challenges are the messy and noisy clinical records, case complexities, along with the huge ICD10 code space. Previous work mainly relied on discharge notes for prediction and was applied to a very limited data scale. We propose an ensemble model incorporating multiple clinical data sources for accurate code predictions. We further propose an assessment mechanism to provide confidence rates in predicted outcomes. Extensive experiments were performed on two new real-world clinical datasets (inpatient & outpatient) with unaltered case-mix distributions from Maharaj Nakorn Chiang Mai Hospital. We obtain multi-label classification accuracies of 0.73 and 0.58 for average precision, 0.56 and 0.35 for F1-scores and 0.71 and 0.4 accuracy in predicting principal diagnosis for inpatient and outpatient datasets respectively.

## 1. Introduction

ICD100 is the World Health Organization's 10[th] version of medical classification diagnosis representing diseases by code. Each code has 7 characters, 1-3 define the category of the disease, 4-6 define the body site and severity, while character 7 is an extension. It is adopted by hospitals, health insurance companies, and public health agencies across the world. Codes assist in easy storage, monitoring, analytics and research [1,2]. The huge number of codes, the diversity and complexity in the structure of medical records greatly increase the difficulty of manual coding. The average human coding accuracy was found to be between 70-75% which is a significant risk [3,6]. Coding errors can have severe impacts on many areas such as billing resulting in underpayments [3,4] and loss of visibility of epidemics/pandemics [5]. Consequently, there is a great demand for automatic and efficient ICD10 coding to bridge this gap.

The problem was previously formulated as multi-class classification (predicting only principal diagnosis) [16], multi-label classification [14,17] and ranking [18]. Individual data sources were employed to predict diagnosis (e.g. physician notes and discharge summaries) [7-10], radiology reports [11,12], medical imaging [13]. Incorporating multiple sources has also been studied [14,15]. On the modeling side, many solutions were proposed such as rule-based replicating human efforts (e.g. coding frequency) [7]. Supervised machine learning was among the most studied techniques [8,9,10,19] compared to unsupervised based models [20]. Deep learning tackled the problem employing different neural networks such as recurrent, convolutional and attention based models [8,10,11,14]. However, most previous work mainly relied on discharge texts. Additionally, these models were trained and evaluated on limited subsets of codes (e.g. ~50 frequent codes) which does not reflect real-life coding complexities.

In this paper, an ensemble model to predict ICD10 codes is presented by incorporating four data types: tabular (admissions), structured (prescribed medications), unstructured (free-text radiology reports), and semi-structured (laboratory tests). Towards this goal, the following contributions were made:

[1] Growing Data yassein, alexander, terence @growingdata.com.au
[2] BioMedical Informatics center, Faculty of Medicine, Chiang Mai University
[3] Department of Surgery, Faculty of Medicine, Chiang Mai University
[4] Department of Family Medicine, Faculty of Medicine, Chiang Mai University
[5] Department of Computer Engineering, Faculty of Engineering, Chiang Mai University
[6] Department of Electrical Engineering, Faculty of Engineering, Chiang Mai University
[7] Biomedical Engineering Institute, Chiang Mai University
8 Maharaj Nakhon Chiang Mai hospital, Chiang Mai, Thailand
9 IT department of Maharaj Nakhon Chiang Mai hospital

- The full exploitation of the knowledge from multiple sources with varied data complexities
- The proposition of an ensemble model to predict ICD10 codes (principal diagnosis & comorbidity)
- The proposition of a confidence assessment network to qualify outcome predictions

To the best of our knowledge, none of the existing research on medical coding consider similar setting of extreme multi-label classification on large-scale real-life case-mix with no discharge notes. Extensive experiments on two real-world datasets quantitively show the superiority of the proposed model.

## 2. Methodology

In this section, the proposed ensemble model is introduced and the problem is formalized as a multi-label classification problem. The model combines evidence from multiple data sources namely: laboratory tests, prescribed medications, radiology reports, and admission data. For each data source, a Feedforward Neural Network is trained to predict the list of ICD10 codes individually. An "Ensemble Network" is placed on top of sub-networks to learn the specialization of each in predicting different diagnosis as shown in Figure 1. It mimics a senior clinical coder given predictions from junior ones, then based on his experience, decides on the correct diagnosis.

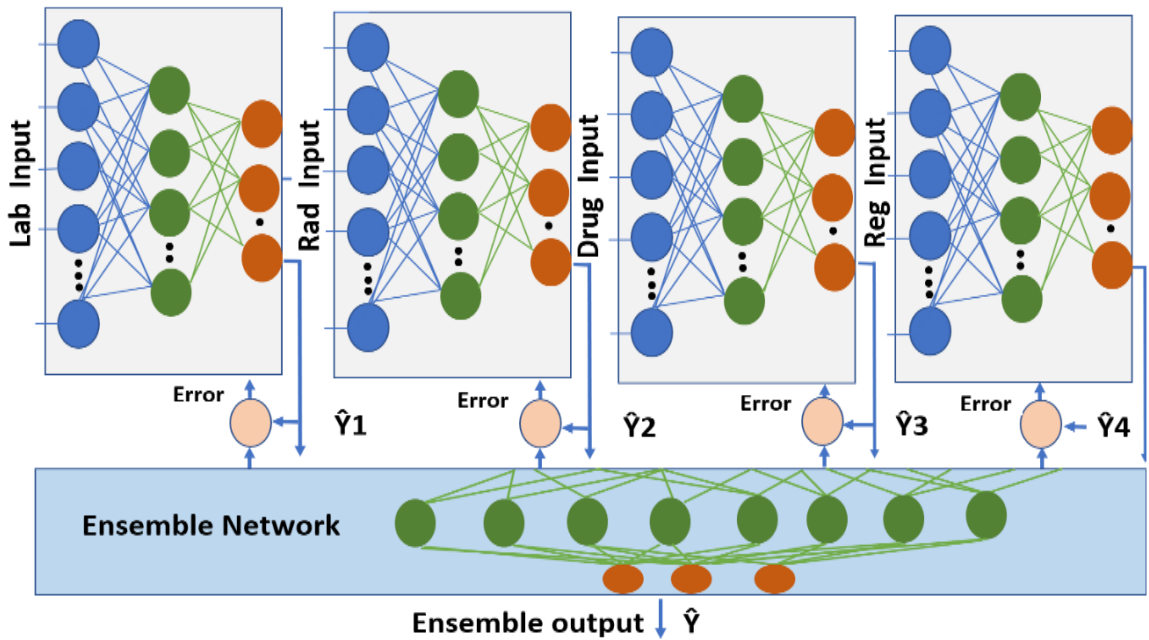

**Figure 1.** ICD10 ensemble model structure

### 2.1 Individual Networks Structure

A feed-forward neural network architecture is adopted for modelling each data source to capture the complex representations from the data. Lab-data network consists of two trainable layers where the hidden layer is activated via a ReLU function. Sigmoid activation function is used for the output layer to predict the multi-label ICD10 codes. The dropout layer is set with a rate of 0.3. The output dimension is set with ICD10 code space size shown in Table 1. Binary cross entropy loss function is chosen to train the model in predicting multi-labels [22]. Radiology network is a 3-layer network with dropout layer of rate of 0.25. Medications network resembles the lab network but with a dropout rate of 0.35.

### 2.2 Ensemble Model Structure

Traditionally, predictions of ensemble models are averaged form multiple sub-models or combined with preset weights. However, in proposed setting, individual networks are trained separately and their predictions (ICD10 labels) are fed as input to the ensemble model. Then, it learns how best to combine and weight the contributions of sub-optimal networks dynamically. Moreover, the model accounts for episodes of care having missing inputs through weight shifting in these cases.

*2.3 Prediction Confidence Network*

At prediction time, users cannot tell how good the result is. Hence, stems a need for a real-time assessment utility for predictions. The proposed confidence network consists of two separate models as shown in Figure 2. First, ICD10 predictor, then prediction errors coupled with original input are coupled to predict confidence score for every record. In training, the model uses all clinical data features plus predicted labels as inputs. It is a regression problem that can be optimized by minimizing MSE. Given unseen test sets, ICD10 codes are predicted, then a discrepancy level is predicted along with a computed threshold to estimate confidence rates. In real life scenario, such system can delegate low-confidence predictions to a human coder automatically. This makes the model viable enough to be commercially adopted.

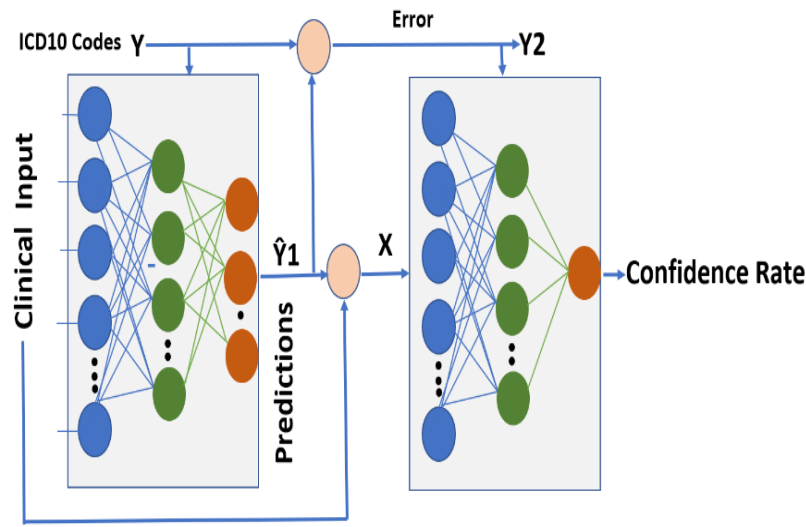

**Figure 2.** Confidence network structure

## 3. Experimental *Setup*

*3.1 Dataset Description*

**Table 1.** Datasets overall statistics

| Dataset | #Records | Avg#codes/ case | Max #codes/case | #Unique codes | #Unique combinations of codes |
|---|---|---|---|---|---|
| Inpatient | 588,932 | 3.43 | 47 | 12,382 | 250k |
| Outpatient | 475,508 | 1.62 | 12 | 9,970 | 80k |

The data is recorded for the period from 2006 to 2019 and Table 1 highlights the overall statistics[11]. Inpatient/Outpatient datasets each contain four sources (medications, laboratory, radiology & admission). Inpatient-medications comprises 5M unique medications with 13.47 average prescriptions per patient. Outpatient comprises 3M unique medications and 3.22 prescriptions on average. Inpatient-radiology has 69.7k records, while outpatient has 98.2k with 1.5k words on average per report. Inpatient-laboratory has 500k records with 15 tests per patient on average, while 400k in the outpatient with an average of 5 tests. Admission data comprises of age, gender, admission & discharge dates.

*3.2 Pre-processing*

---

[11] *Hospital name will be released after review*

Very rare cases with fewer than 3 examples are filtered. Prediction of each ICD10 code is limited to first 3 letters (reflecting disease category). Radiology reports are cleaned by removing special characters & stop words, transformed to lower case and split into 5 sentences. Sentence embeddings are extracted for each using BioSentVec medical embeddings [23]. Test location (e.g. hand, brain) and position (e.g. hand PA) are transformed from text to binarized representation.

The laboratory dataset is the most challenging with raw data of 66 GB in size with multiple formats including text, numeric and mixtures. First, data is cleaned by unifying casing and removing bad characters using one of three parsing modes: numeric, categorical and mixture (e.g. neg, -ve, positive, 1+, 716.3 iu/ml, range 2-4). Next, raw values are transformed into logarithmic scale to remove skewness, then normalised to values [0 -1]. Finally, for each patient, tests are represented by 3 values: the number times a test is performed and the minimum and maximum recorded test values.

The only preprocessing for the medications dataset is to remove cancelled prescriptions and to binarize into multi-label sparse vectors.

For admission-data, gender is encoded with 0 or 1, while age is binned into age groups. Adding length of stay to encode the severity of inpatient cases. To incorporate seasonality of diseases, admission day, month and year are encoded.

*3.3 Evaluation Measures*

Assessing performance for multi-class-multi-label classification is more about which combinations of labels are right than exact sets. The following classification evaluation measures are adopted [24]. F1 score: harmonic mean of precision and recall with micro-averages. Jaccard similarity: measure of similarity distance. Average precision: combines recall and precision for ranked sets. Coverage error: bias error when target population does not coincide with sampled population. Ranking loss: average number of incorrect label pairs weighted by the set label size and the number of missing labels. Accuracy: (#correct/#records) only for principal diagnosis.

*3.4 Experimental Setting*

To reproduce one's experiments, all data was stratified and split into (train-dev-test) sets of (70%-10%-20%). Adam optimizer was used in training on batch sizes of 256. For inpatient, the number of hidden units for medications, lab, radiology and registration are 600, 350, 600 & 30, while input dimensions are 4986, 768, 4563 & 59 respectively. For outpatient, number of hidden units for medications, lab, radiology and registration are 500, 250, 550, 25, while input dimensions are 3008, 474, 3786 & 25 respectively. Python deep learning Keras library was employed for implementation. Pre-processing and training were performed on a 2-GPU (GTX 1060 6GB) machine.

*4. Results & Discussion*

*4.1 ICD10 Multi-label Classification*

Related work to this study mainly relies on discharge notes (descriptions of reached diagnosis) which are not part of this study. Thus, one's model is compared to two very common modelling strategies. First, *"Combined Model"*, where all features from all data sources are stacked to form one vector and trained using same structure as individual networks. Second, *"Averaging Model"*, where sperate networks is trained on individual data sources then their predictions are averaged. In Table 2, one's model performs consistently better on all evaluation metrics compared to other baselines. Improvement of 4-5% was witnessed on inpatient and 2-3% on outpatient datasets. Moreover, it was found that the accuracy of some sources (e.g. Medications) are often comparable to that of combined sources. This is because each source converges at different rates with different model complexities. Thus, stacking of features results in some features from one data source dominate the feature space. Complex modalities such as laboratory degrade the overall accuracy because of its slower convergence rate. Thus, leading to the under-exploration of other modalities. This observation is supported by [21], stating that some deep networks can quickly find an optimal minimum from some modalities faster than others. That is why learning one's model each modality separately to fully encodes the variability data complexity levels. Hence, the expertise of each network can be successfully learned leading to combining the right evidence for the best prediction. Additionally, high prediction accuracy for principal diagnosis was also witnessed. This is great news for medical billing applications relying essentially on its identification.

**Table 2.** ICD10 Classification accuracy

| Dataset | Model | Average Precision | Ranking Loss | Coverage Error | Jaccard Similarity | F1 | Accuracy (Principal) |
|---|---|---|---|---|---|---|---|
| Inpatient | Combined | 0.69 | 1.66 | 25.81 | 0.46 | 0.52 | 67% |
| | Averaging | 0.67 | 1.72 | 25.92 | 0.14 | 0.18 | 29% |
| | Ensemble | **0.73** | **0.77** | **12.88** | **0.50** | **0.56** | **71%** |
| Outpatient | Combined | 057 | 1.90 | 26.50 | 0.32 | 0.33 | 38% |
| | Averaging | 0.55 | 3.64 | 36.33 | 0.04 | 0.04 | 6% |
| | Ensemble | **0.58** | **1.88** | **26.20** | **0.33** | **0.35** | **40%** |

In Table 3, top 5 disease categories ranked by accuracy are shown. The top categories of inpatient dataset is on average higher than 80%. For outpatient, it is lower, one reason is the lack of historical traces of follow-up visits which can be essential for chronic diseases. For instance, the accuracy for neoplasms is 38% and ranked 9[th] for outpatients compared to 5[th] rank for inpatients where there is usually more evidence (e.g. more tests, length of stay).

**Table 3.** Top 5 diagnosis categories prediction accuracy

| Outpatient Dataset | | Inpatient Dataset | |
|---|---|---|---|
| Diagnosis Category | Accuracy | Diagnosis Category | Accuracy |
| Endocrine, nutritional & metabolic diseases | 71% | Conditions originating in the perinatal period | 97% |
| Diseases of the circulatory system | 68% | Pregnancy, childbirth and the puerperium | 91% |
| Diseases of the blood and immune disorders | 64% | Factors influencing health status | 90% |
| Certain infectious and parasitic diseases | 64% | Diseases of the eye and adnexa | 81% |
| Diseases of the genitourinary system | 58% | Neoplasms | 80% |

*4.2 Ablation Study*

In Table 4, the effect of removing sub-networks from our architecture is shown to gain better understanding of the modelling behaviour. It is evident that prescriptions' data show the best individual performance. This can be attributed to the fact that prescription usually occurs later in the episode of care and better associated with diagnosis. Laboratory tests were found to be more predictive than radiology due to easier learning from numeric evidence. Combing the three sources was found to be essential in untangling complex cases yielding 4% prediction accuracy improvement on F1 & Jaccard similarity and 5% for principal diagnosis. Combining admission and seasonality add only 1 % to improvements as shown in Table 2.

**Table 4.** ICD10 Ablation study classification accuracy

| Dataset | Model | Average Precision | Ranking Loss | Coverage Error | Jaccard Similarity | F1 | Accuracy (Principal) |
|---|---|---|---|---|---|---|---|
| Inpatient | Medications | 0.71 | 0.90 | 14.9 | 0.46 | 0.52 | 67% |
| | Lab | 0.37 | 4.78 | 70.5 | 0.13 | 0.16 | 23% |
| | Rad | 0.13 | 36.2 | 481. | 0.04 | 0.05 | 8% |
| | Medications +Lab | 0.71 | 0.88 | 14.7 | 0.47 | 0.53 | 68% |
| | Medications +Lab+Rad | 0.72 | 0.76 | 12.7 | 0.50 | 0.56 | 70% |
| Outpatient | Medications | 0.57 | 1.96 | 27.1 | 0.30 | 0.32 | 36% |
| | Lab | 0.22 | 8.17 | 108 | 0.05 | 0.06 | 8% |
| | Rad | 0.30 | 8.61 | 113 | 0.09 | 0.13 | 12% |
| | Medications +Lab | 0.57 | 1.88 | 25.6 | 0.32 | 0.34 | 38% |
| | Medications +Lab+Rad | 0.58 | 1.67 | 23.3 | 0.34 | 0.36 | 40% |

*4.3 Confidence Evaluation Experiments*

In Table 5, the accuracy at different confidence thresholds is shown. Given input evidence, confidence levels can be speculated in one's predictions without ground-truth knowledge. For example, one's model shows a verified confidence of 85% for 50% of the predictions. This entails deeper level of understanding of the various combinations of clinical input distributions which can verify one's predictions accuracy.

**Table 5.** Prediction confidence and scope for inpatient dataset

| Data Scope | AP | CR | RL |
|---|---|---|---|
| 3% | 0.97 | 12.3 | 0.270 |
| 10% | 0.95 | 15.2 | 0.272 |
| 20% | 0.92 | 16.2 | 0.313 |
| 30% | 0.90 | 20.1 | 0.367 |
| 50% | 0.85 | 24.1 | 0.546 |

## 5. Conclusion

A novel ensemble deep learning model is proposed to predict ICD10 codes harnessing knowledge from multiple clinical data sources. Multiple data modalities were consumed including unstructured, semi-structured and structured tabular data. Due to the sensitivity of miscoding, an automatic confidence assessment model is further proposed for proven trustworthy prediction results. Extensive experiments on two new real-life datasets show the superiority of one's model to be adopted in real clinical coding practice in pre and/or post discharge setting. The positive results have opened the door to further future work. One's models can be applied in real time coding by continuously learning and updating the model with new evidence on the fly in production.